# Magnetic Resonance Connectome Automated Pipeline

William R. Gray, *Member, IEEE,* John A. Bogovic, *Student Member, IEEE,* Joshua T. Vogelstein, Bennett A. Landman, *Member, IEEE*, Jerry L. Prince, *Fellow, IEEE,* R. Jacob Vogelstein

*Abstract*— This manuscript presents a novel, tightly integrated pipeline for estimating a connectome, which is a comprehensive description of the neural circuits in the brain. The pipeline utilizes magnetic resonance imaging (MRI) data to produce a high-level estimate of the structural connectivity in the human brain. The Magnetic Resonance Connectome Automated Pipeline (MRCAP) is efficient and its modular construction allows researchers to modify algorithms to meet their specific requirements. The pipeline has been validated and over 200 connectomes have been processed and analyzed to date.

This tool enables the prediction and assessment of various cognitive covariates, and this research is applicable to a variety of domains and applications. MRCAP will enable MR connectomes to be rapidly generated to ultimately help spur discoveries about the structure and function of the human brain.

## I. INTRODUCTION

A connectome is a comprehensive description of the neural network in a brain [1]. Depending on the application and the available data, the nodes of the network may correspond to individual neurons, or large collections thereof [2]. In living humans, connectomes can be estimated from multimodal magnetic resonance imaging (MRI), which provides detailed information about brain connectivity at millimeter resolution. These MR connectomes have been postulated to be able to predict and assess mental properties such as intelligence, psychopathy, and dementia [1], [3-5]. Indeed, the implications of connectomics are wide-ranging, with applications for medicine, education, the military and intelligence communities, computer science, pattern recognition, and many other fields.

During the past few years, there have been a number of research groups working on methods and techniques for generating MR connectomes. This complex task requires tools such as FSL/Freesurfer [6], [7], MedINRIA [8], or BrainVISA [9] that combine cortical labeling and segmentation with diffusion tensor imaging. Custom software scripts are typically developed to link the necessary routines and obtain the required connectivity measurements (though see [10]).

In this article we present a new automated method to obtain MR connectomes. Our pipeline generates connectomes through an integrated graphical programming environment, which offers a tightly coupled set of software routines that are available for multiple platforms and input data types. The pipeline is based on the Java Image Science Toolkit (JIST) [11], [12], which runs in conjunction with the Medical Image Processing, Analysis and Visualization (MIPAV) software [13].

## II. OVERVIEW

Our MR connectome automated pipeline (MRCAP) combines diffusion-weighted images with structural MR images to generate an MR connectome derived from connectivity measurements between anatomically-defined cortical regions. The connectome is represented by a connectivity matrix suitable for input to graph theoretic [14], [15] or statistical [5], [16-20] algorithms which can infer meaning from the data. The latest stable release of MRCAP is available for download from NITRC at www.nitrc.org/projects/mrcap, and the code base is actively being developed. Routines (called modules) can be updated and replaced as needed to meet the requirements of the individual researcher. Indeed, because there is no scientific consensus on how to best estimate synaptic connectivity from MRI data, the pipeline is designed to serve as a testbed to explore different approaches.

MRCAP is comprised of three JIST layouts—structural, diffusion, and connectivity—each of which consists of a collection of modules (see Figure 1). The modules themselves

Manuscript received April 25, 2011. This work was supported by the NSA Research Program on Applied Neuroscience and NIH/NINDS 5R01NS056307.

W. R. Gray is with the Johns Hopkins University, Baltimore, MD 21210 USA and the Johns Hopkins University Applied Physics Laboratory, Laurel, MD 20723, USA. (phone: 443-778-9538; fax: 443-778-5342; email: willgray@jhu.edu).

J. A. Bogovic, J. T. Vogelstein, and J. L. Prince are with the Johns Hopkins University, Baltimore, MD 21210, USA (email: {bogovic, joshuav, prince} @ jhu.edu).

B. A. Landman is with Vanderbilt University, Nashville, TN 37235, USA (email: bennett.landman@vanderbilt.edu).

R. J. Vogelstein is with the Johns Hopkins University, Baltimore, MD 21210 USA and the Johns Hopkins University Applied Physics Laboratory, Laurel, MD 20723, USA (email: jacob.vogelstein@jhuapl.edu).

are assembled from a variety of algorithms, authors, and methods; our contribution is the integration of these tools into one automated processing flow. The structural layout operates on the structural MR data and performs skull stripping and parcellation of the brain into labeled gyral regions. The diffusion layout estimates tensors, computes fractional anisotropy (FA) values, and performs fiber tracing. Finally, the connectivity layout registers the diffusion data to the structural space, combines the putative fiber tracts with the associated gyral region labels, and uses the FA information to generate a connectome (in the form of a mathematically-convenient adjacency matrix). Alternate measures of connection strength (such as fiber count and fiber length) are also available. A more detailed explanation of the pipeline functionality is described below.

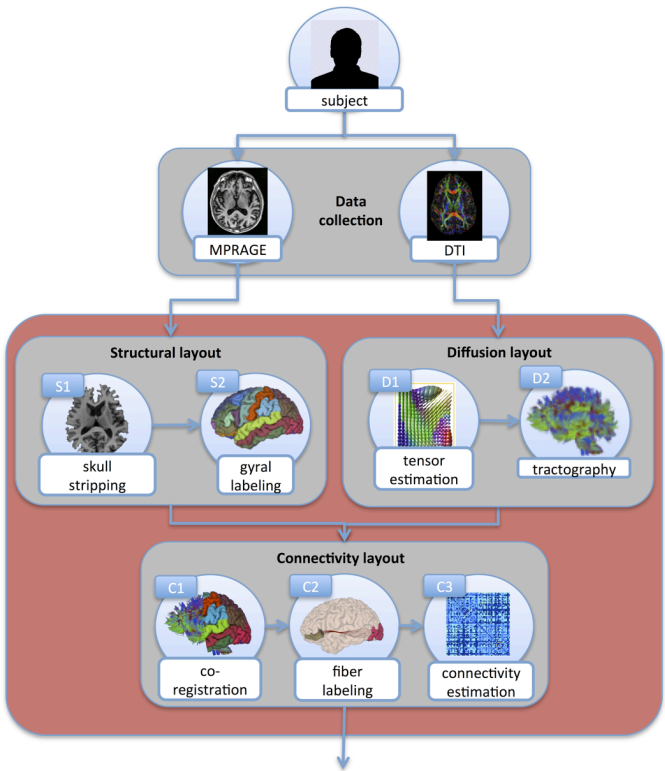

Figure 1: Overall pipeline processing flow: this figure gives a high-level overview of the structural, diffusion and connectivity processing layouts. The layout uses magnetization-prepared rapid acquisition with gradient echo (MPRAGE) and diffusion-weighted magnetic resonance imaging (DW-MRI) scans to estimate a connectome.

### III. METHODS

The JIST environment allows a researcher to graphically design a processing pipeline that generates a variety of useful outputs for validation and further analysis. It also provides an Application Programming Interface (API) that facilitates the interoperability of modules developed by various authors. Graphical tools allow for easily adding, swapping, and/or modifying modules as needed. Because it is based on the Java programming language, JIST can run on many computer architectures [12]. Figure 2 shows a screenshot of the JIST programming environment and our pipeline. In the JIST window, the processing steps are represented by input file modules (blue), algorithm modules (red), and lines connecting inputs and outputs (black). Within the JIST framework, researchers can select from a variety of pre-defined modules and assign processing parameters. Both JIST and MIPAV are freely available for download at www.nitrc.org/projects/jist and www.nitrc.org/projects/mipav, respectively.

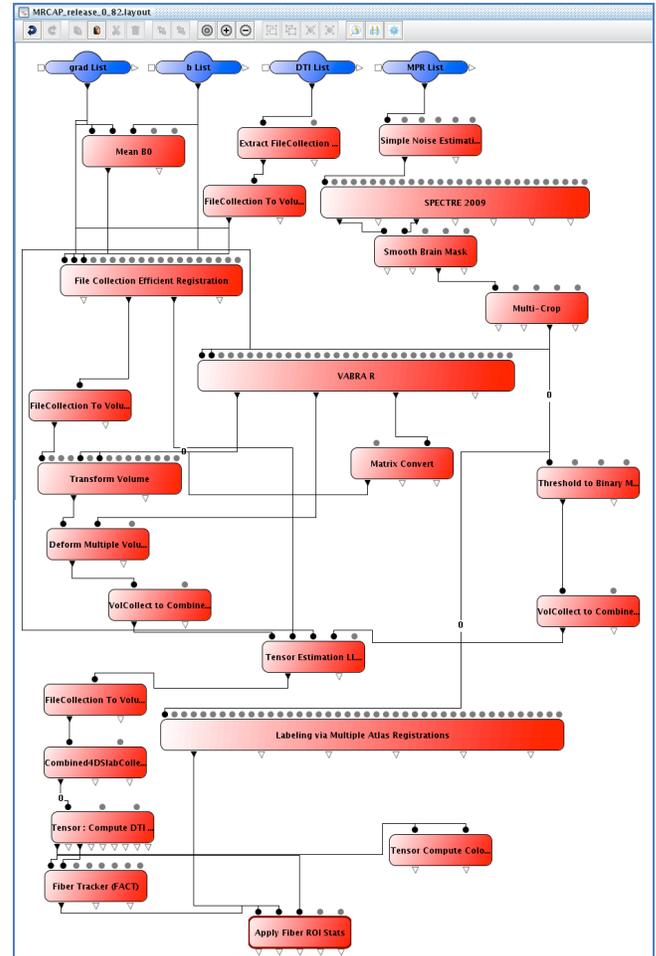

Figure 2: An overview of the Magnetic Resonance Connectome Automated Pipeline (MRCAP) in the Java Image Science Toolkit (JIST) framework.

#### A. Input Data

The pipeline accepts the diffusion and structural MR data from a subject, the associated metadata, and user-specified parameters as inputs. Various input image formats are supported through built-in JIST modules, including XML, PAR/REC, NIFTI, and DICOM.

#### B. Structural Processing

**S1:** Structural image processing begins with the SPECTRE algorithm [21], which removes the skull and other non-brain tissue using a joint registration and tissue classification technique. The tissue classification is performed using FANTASM, a robust fuzzy C-means intensity classification algorithm [22]. This allows for the identification of high-intensity skin and adipose tissue and low-intensity bone matter, all of which can be subsequently eliminated. This result is smoothed and is used as an input for diffusion image co-registration in the connectivity layout.

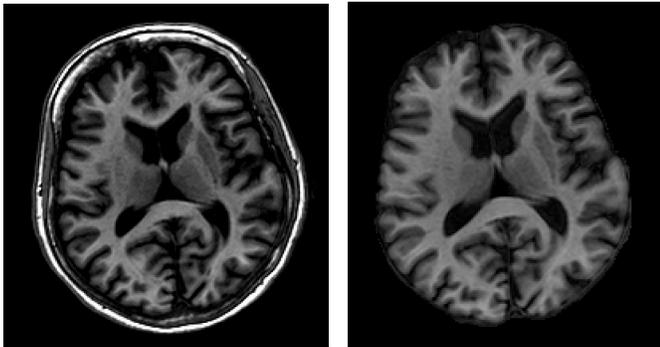

Figure 3: Axial slice of original MPRAGE image and corresponding skull stripped result

**S2:** In the second step of the structural processing layout, the brain is divided into a set of 70 regions defined by the Desikan gyral label atlas [23]. Parcellation is achieved by registering one or more template brains to the subject brain using VABRA, a vectorized form of the Adaptive Bases registration Algorithm (ABA) [24]. This algorithm performs nonrigid intensity-based registration using normalized mutual information as a cost function and models the deformation as a linear combination of radial basis functions. The results from the different template registrations are subsequently combined using STAPLE [25].

### C. Diffusion Processing

**D1:** Using the diffusion-weighted images, the "diffusion tensor" is estimated for each voxel using a log-linear minimum mean squared error measure [26]. In the context of our pipeline, a diffusion tensor is a local model of the diffusion of water in the brain; this measurement is influenced by tissue microstructure, particularly axonal projections. These tensors enable the computation of fractional anisotropy (FA), a scalar value derived from the tensor that roughly describes the directionality of diffusion and can indicate the "coherency" of axonal bundles summarized by the tensor.

**D2:** From the computed tensors, streamlines are derived with the FACT algorithm [27], a fast, deterministic algorithm for reconstructing fibers. FACT is a classical method that has been shown to recover many important fiber tracts [28], and is widely applied in the neuroscience community despite its inability to resolve crossing fibers. If desired, probabilistic tractography or other algorithms may be implemented instead [6], [29], [30]. Each computed fiber tract represents the estimated location of a large group of axons, which are signaling pathways (i.e., connections) between brain regions.

### D. Connectivity Processing

**C1:** At the beginning of the connectivity layout, the diffusion image data is preprocessed and co-registered to the structural output data using VABRA.

**C2:** Each fiber streamline traverses a (potentially large) number of voxels. We postulate that axonal fibers exist and connect any pair of voxels that a streamline traverses; therefore two regions containing a voxel within a fiber streamline are assumed to be connected (see Figure 4).

**C3:** To obtain an estimate of connection strength, the FA values associated with the voxels that a fiber passes through are averaged to create a mean FA value for that fiber. The overall measure of connectivity between any two regions is the average of the mean FAs for all fibers connecting those two regions.

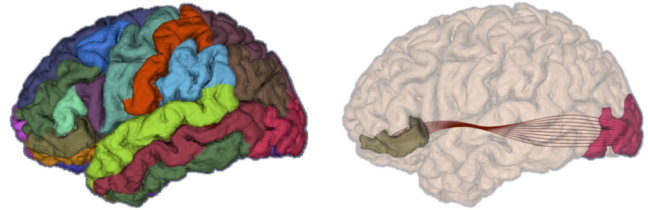

Figure 4: The left image contains an illustration of a subject's delineated gyral regions. The right image shows a notional example of the fiber streamlines connecting two of those regions.

Because we divide each brain into 70 regions, our MR connectomes are theoretically characterized by 70 x 70 = 4,900 values. However, because we cannot assign a polarity to a streamline, the connections are undirected, implying that the 70 x 70 matrix is symmetric. Furthermore, we do not compute connections within a region, implying that the matrix is hollow (i.e., the diagonal is empty). For these reasons, the final dimensionality of the output is 2,415, representing the connection strength between each of the 2,415 pairs of cortical regions. An example of the MR connectome result for one subject, represented by a connectivity matrix, is shown in Figure 5. Every region pair has an associated edge with a mean FA value between zero and one, with red colors representing stronger connections. Node pairs exhibiting no connectivity are assigned a value of zero, shown as dark blue in the figure. An alternative three-dimensional view of selected cortical connections is shown in Figure 6.

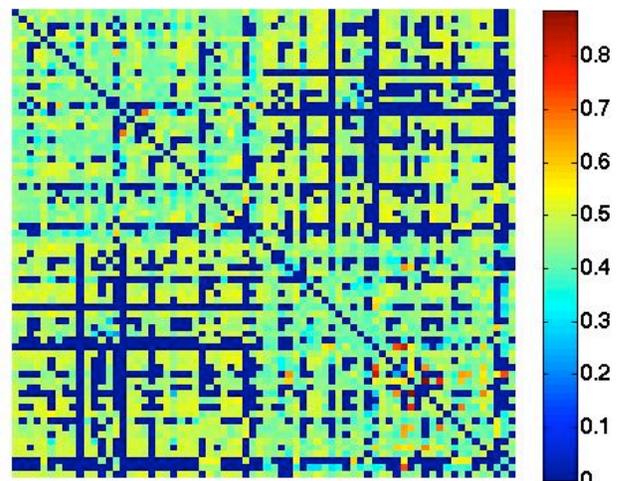

Figure 5: Example MR Connectome, showing the mean(mean FA) value for all of the fibers between each pair of gyral regions.

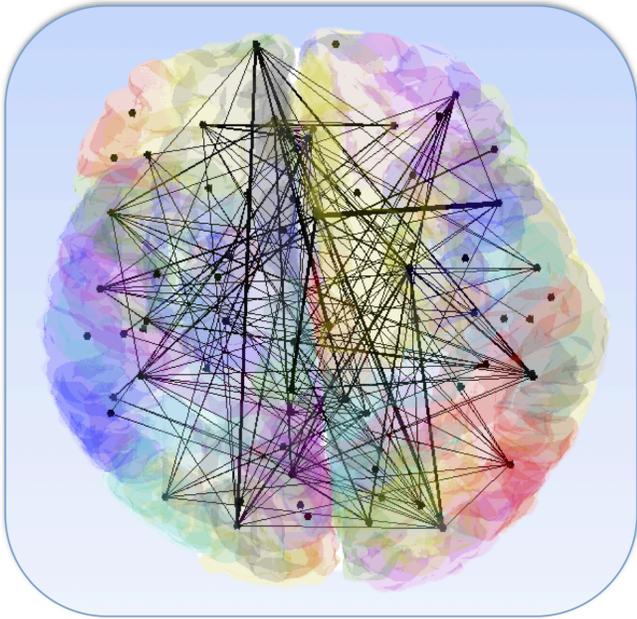

Figure 6: This figure superimposes the subject's connectome on a parcellated brain. The various colors indicate different gyral regions, and the black dots illustrate region centroids. The lines between regions indicate connectivity, and the line weights show connection strength. For clarity only the strongest 15% of connections are shown, based on mean fractional anisotropy.

## IV. Discussion

We are currently using the MR connectomes generated by MRCAP to study gender differences and memory and motor tasks in normal aging. The end-to-end pipeline has been successfully employed to automatically process more than 200 datasets. It takes approximately eight hours to generate a connectome for a subject using a single-core 3 GHz CPU and 10 GB RAM. We execute MRCAP on a high-performance cluster, and in this hardware configuration we can run one subject on each core in parallel.

To verify that the connectivity measures produced by MRCAP are meaningful, we have validated the results using a variety of qualitative and quantitative techniques. For example, we developed tools to visualize fiber bundles and ensure that the streamline paths appear to have reasonable trajectories (Figure 7). Additionally, selected inter- and intra-hemispheric connections were compared and evaluated for symmetry. Finally, connectivity matrices based on both FA connectivity values and fiber counts (two alternative measures of connection "strength") were investigated both visually and quantitatively to look for underlying processing defects.

The underlying resolution of our data is approximately 1-2 millimeters per voxel; current MRI resolution is limited to this order of magnitude. We assert that this resolution is sufficient to measure macroscopic information pathways in the brain to obtain a high-level assessment of the connectivity between anatomical brain regions. This MRI-based approach is attractive because it allows measurements to be taken *in vivo*, using standard imaging modalities and acquisition sequences.

Connectome inference in humans may be possible at higher (nanometer) resolutions, but the necessary data acquisition and analysis protocols are still in development and current methods require imaging of *ex-vivo* tissue [31].

Connectomes have applications to many areas of research. This includes investigations into understanding brain plasticity and recovery of function following central nervous system injury. From a healthcare perspective, this could be used to predict disease susceptibility or likely disease progression, and could suggest individually tailored treatments. In a military context, this could help predict soldier vulnerability to combat effects (e.g., post-traumatic stress disorder), and could also be used to augment staff selection methods and learning assessments for a variety of purposes [32].

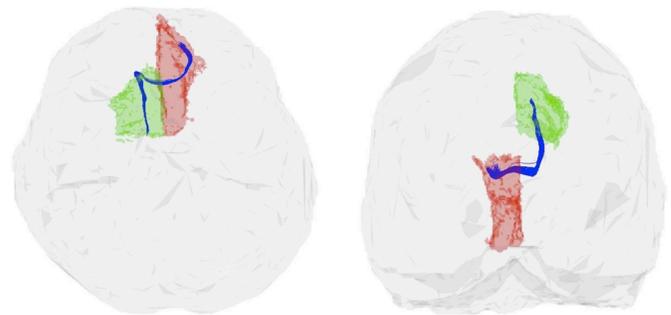

Figure 7: Fiber streamlines visualized in axial (left) and coronal (right) views. This is an anecdotal result showing the fiber streamlines (shown in blue) that connect two gyral regions in the brain (shown in red and green). This information will be used to assess the connection strength between these regions, and the result will be used to populate a single entry in the final connectivity matrix.

## V. Conclusion

In summary, the MRCAP pipeline allows for automatic assessment of the connectivity of a human brain in a robust, flexible, graphical environment (with a command-line interface for batch processing). Processing is efficient and intermediate outputs allow for validation and provide documentation of results. We hope that this tool will enable MR connectomes to be rapidly generated for many subjects and, ultimately, spur discoveries about the structure and function of the human brain.